\documentclass[aps,prl,twocolumn,superscriptaddress,showpacs]{revtex4}

\begin{document}


\title{Constraints on Nucleon Decay via ``Invisible'' Modes \\
from the Sudbury Neutrino Observatory}



%
\newcommand{\ubc}{Department of Physics and Astronomy, University of 
British Columbia, Vancouver, BC V6T 1Z1 Canada}
\newcommand{\bnl}{Chemistry Department, Brookhaven National 
Laboratory,  Upton, NY 11973-5000}
\newcommand{\carleton}{Ottawa-Carleton Institute for Physics, Department of Physics, Carleton University, Ottawa, Ontario K1S 5B6 Canada}
\newcommand{\uog}{Physics Department, University of Guelph,  
Guelph, Ontario N1G 2W1 Canada}
\newcommand{\lu}{Department of Physics and Astronomy, Laurentian 
University, Sudbury, Ontario P3E 2C6 Canada}
\newcommand{\lbnl}{Institute for Nuclear and Particle Astrophysics and 
Nuclear Science Division, Lawrence Berkeley National Laboratory, Berkeley, CA 94720}
\newcommand{\lanl}{Los Alamos National Laboratory, Los Alamos, NM 87545}
\newcommand{\oxford}{Department of Physics, University of Oxford, 
Denys Wilkinson Building, Keble Road, Oxford, OX1 3RH, UK}
\newcommand{\penn}{Department of Physics and Astronomy, University of 
Pennsylvania, Philadelphia, PA 19104-6396}
\newcommand{\queens}{Department of Physics, Queen's University, 
Kingston, Ontario K7L 3N6 Canada}
\newcommand{\uw}{Center for Experimental Nuclear Physics and Astrophysics, 
and Department of Physics, University of Washington, Seattle, WA 98195}
\newcommand{\triumf}{TRIUMF, 4004 Wesbrook Mall, Vancouver, BC V6T 2A3, Canada}
\newcommand{\ral}{Rutherford Appleton Laboratory, Chilton, Didcot, 
Oxon, OX11 0QX, UK}
\newcommand{\uta}{Department of Physics, University of Texas at Austin,
Austin, TX 78712-0264}
\newcommand{\iusb}{Department of Physics and Astronomy, Indiana University, South Bend, IN}

\affiliation{	\ubc	}
\affiliation{	\bnl	}
\affiliation{	\carleton	}
\affiliation{	\uog	}
\affiliation{	\lu	}
\affiliation{	\lbnl	}
\affiliation{	\lanl	}
\affiliation{	\oxford	}
\affiliation{	\penn	}
\affiliation{	\queens	}
\affiliation{	\ral	}
\affiliation{	\uta	}
\affiliation{	\triumf	}
\affiliation{	\uw	}

\author{	S.N.~Ahmed	}\affiliation{	\queens	}									
\author{	A.E.~Anthony	}\affiliation{	\uta	}									
\author{	E.W.~Beier	}\affiliation{	\penn	}									
\author{	A.~Bellerive	}\affiliation{	\carleton	}									
\author{	S.D.~Biller	}\affiliation{	\oxford	}									
\author{	J.~Boger	}\affiliation{	\bnl	}									
\author{	M.G.~Boulay	}\affiliation{	\lanl	}									
\author{	M.G.~Bowler	}\affiliation{	\oxford	}									
\author{	T.J.~Bowles	}\affiliation{	\lanl	}									
\author{	S.J.~Brice	}\affiliation{	\lanl	}									
% ...
			
\collaboration{SNO Collaboration}
\noaffiliation

\affiliation{}


\date{\today}

\begin{abstract}
Data from the Sudbury Neutrino Observatory have been used to constrain
the lifetime for nucleon decay to ``invisible'' modes, such as 
$n \rightarrow 3 \nu$. The analysis was based on a search for $\gamma$-rays
from the de-excitation of the residual nucleus that would result from the 
disappearance of either a proton or neutron from $^{16}$O.  
A limit of $\tau_{\rm inv} > 2\times10^{29}$ years is 
obtained at 90\% confidence for either neutron or proton decay modes. This
is about an order of magnitude more stringent than previous constraints
on invisible proton decay modes and 400 times more stringent 
than similar neutron modes.
\end{abstract}


\maketitle


Experimental signatures of the grand unification of the electroweak
and strong interactions have been sought with
increasing sensitivity for the past twenty-five years.  Much
effort has gone to identifying specific decay modes of free protons
and bound nucleons as signatures of grand unification, but no signal
has been observed to date. Decay modes more unusual than
those typically explored cannot, however, be ruled out (see, for example, \cite{Pati}).
A recent paper has even suggested a model in which 
$n \rightarrow 3 \nu$ becomes the dominant mode \cite{Mohapatra}. Thus, 
the search for any mode
which may have been missed by previous experiments is of fundamental interest.
In this paper, the Sudbury Neutrino Observatory (SNO) \cite{boger} is used 
to search for what we will refer to as ``invisible'' decay channels,
i.e., those in which no visible energy is deposited in the detector 
via direct production of energetic, charged particles.

The search utilizes SNO's unique detection capabilities for
low energy $\gamma$-rays, based on the Cherenkov light produced by 
the resulting Compton-scattered electron. A
generic signature for nucleon disappearance in $^{16}$O arises from the 
subsequent de-excitation of the residual nucleus [4,5].  
Approximately 45\% of the time, the de-excitation of either 
$^{15}$O$^*$ or $^{15}$N$^*$ results in the production of a $\gamma$-ray
of energy 6-7 MeV.  SNO detects these $\gamma$-rays with
good efficiency.  In fact, the primary energy calibration source used
by SNO is the 6.13-MeV $\gamma$-ray produced in the de-excitation of
$^{16}$N.

A background to this nucleon-decay signal in SNO results from neutral 
current (NC) interactions of solar neutrinos. This is due to the 
$\gamma$-rays of similar
energies that are produced as a result of neutron captures on nuclei 
in the detector. In Phase I (D$_2$O), neutrons were detected through 
observation of the single 6.25-MeV $\gamma$-ray resulting from neutron 
capture on deuterium. This capture efficiency is 0.29, but the
threshold applied to the analysis to limit low energy backgrounds
reduces the overall neutron detection efficiency to $\epsilon_n =
0.144\pm0.005$ \cite{nc_prl}. In Phase II (D$_2$O+NaCl), two tonnes 
of NaCl were added to the one kilotonne of D$_2$O. Neutron captures 
on $^{35}$Cl release 8.6 MeV of energy in $\gamma$-rays, with most capture 
events producing multiple $\gamma$-rays. The corresponding capture 
efficiency in Phase II is 0.90 and,  due to a relatively high analysis 
threshold, the overall neutron detection efficiency is 
$\epsilon_n' = 0.399\pm0.010$ \cite{salt_prl}. The multiple $\gamma$-rays 
from neutron captures in Phase II result in a 
more isotropic distribution of Cherenkov light, which can be used as a
further discriminant for identifying the neutron-induced component.
However, the principal advantage in comparing Phase I and Phase II
data lies in the fact that $\gamma$-rays from the nucleon-decay signal 
are detected with similar efficiencies in SNO, while neutrons produced 
by $^8$B solar neutrinos are detected with very different efficiencies.  
These characteristics are used in what follows to measure an upper limit 
for nucleon disappearance.

  In terms of the data from Phase I of SNO, the rate of nuclear $\gamma$-ray 
production can be related to the apparent production rate of neutrons by
taking account of the detection efficiencies for neutrons, $\gamma$-rays 
and particle misidentification as follows:
  
 \[R_\gamma \epsilon_\gamma f_{\gamma n} = R_n - \epsilon_n \mathcal{R}_{NC} \]
 
\noindent where $R_\gamma$ is the rate of nuclear $\gamma$-ray production due to nucleon 
decay; $\epsilon_\gamma$ is the efficiency for detecting the nuclear
$\gamma$-rays above the analysis energy threshold;
$f_{\gamma n}$ is the fraction of the detected nuclear $\gamma$-rays which are
mistaken for neutrons; $R_n$ is the extracted neutron detection rate nominally 
attributed to NC interactions; $\epsilon_n$ is the neutron detection efficiency
for the fiducial volume and analysis energy threshold; and $\mathcal{R}_{NC}$ is the 
actual production rate of neutrons due to solar neutrino NC interactions. Similarly,
for Phase II data,

 \[R_\gamma{\epsilon^\prime_\gamma} f^\prime_{\gamma n} = R^\prime_n - \epsilon^\prime_n \mathcal{R}_{NC} \]

\noindent Thus,

\[R_\gamma = \frac{R_n - \frac{\epsilon_n}{\epsilon^\prime_n}R^\prime_n}
{\epsilon_\gamma f_{\gamma n} - \epsilon^\prime_\gamma f^\prime_{\gamma n} \frac{\epsilon_n}{\epsilon^\prime_n}} 
\equiv \frac{\Delta R_n}
{\epsilon_\gamma f_{\gamma n} - \epsilon^\prime_\gamma f^\prime_{\gamma n} \frac{\epsilon_n}{\epsilon^\prime_n}}  \]

\noindent where $\Delta R_n$ is the difference between the extracted 
neutron detection rate attributed to NC interactions in Phase I and
that implied by data from Phase II. 

In order to compare Phase I and Phase II rates under the same assumption for the underlying
CC spectrum, results from SNO data were used in which the CC component was constrained
to follow the shape of a standard $^8$B energy spectrum 
\cite{foot}.
Table I summarizes
the relevant results from these two phases. The extracted numbers of CC, NC
and ES (elastic scattering) events include the subtraction of all known
backgrounds (as detailed in \cite{nc_prl} and \cite{salt_prl}), including atmospheric 
neutrino interactions which might identically mimic the nucleon-decay signal via 
the removal of a proton or neutron from $^{16}$O.

\begin{table*}
\begin{flushleft}
\caption{Signal extraction results for CC constrained to $^8$B shape.
Error bars are the quadrature sum of statistical and systematic uncertainties.}
\begin{ruledtabular}
\begin{tabular}{|l||l|l|}
Analysis Parameter  &  Phase I (pure D$_2$O) & Phase II$^*$ (D$_2$O + NaCl) \\
\hline

Fiducial Volume            & $6.97\times10^{8}$cm$^3$ & $6.97\times10^{8}$cm$^3$ \\
Energy Threshold           & $T_{\rm eff}>5$ MeV & $T_{\rm eff}>5.5$ MeV \\
Livetime                   & 306.4 days & 254.2 days \\
CC Events   & 1967.7$\pm$117.9 & 1430.3$\pm$97.1\\
ES Events   &  263.6$\pm$29.2 &  163.7$\pm$23.8 \\
NC Events   &  576.5$\pm$70.4 & 1265.8$\pm$95.2 \\
Neutron Detection Efficiency  & 0.144$\pm$0.005 & 0.399$\pm$0.010 \\
NC Event Rate ($R_n$ \& $R^\prime_n$) 
                & 686.8$\pm$83.9 yr$^{-1}$ & 1817.6$\pm$136.6 yr$^{-1}$\\
Equivalent Phase I NC Rate 
($R_n$ \& $\frac{\epsilon_n}{\epsilon^\prime_n}R^\prime_n$) 
                & 686.8$\pm$83.9 yr$^{-1}$ & 656.0$\pm$49.3 yr$^{-1}$\\

\end{tabular}
\end{ruledtabular}
\noindent {$^*$ The Phase II data set used for this analysis is identical to that 
presented in \cite{salt_prl}}
\end{flushleft}
\end{table*}

The number of NC events extracted in Phase I \cite{nc_prl}
was $576.5$ for a livetime of 306.4 days, yielding a rate of 
$R_n = 686.8\pm83.9$ per year, with statistical and systematic 
uncertainties added in quadrature. Similarly, for Phase II,
$1265.8$ NC events were implied based on a livetime of 254.2 days,
yielding a rate of $R^\prime_n = 1817.6 \pm 136.6 $ per year.
Thus, accounting for the relative neutron detection efficiencies, 
\begin{eqnarray*}
\Delta R_n & = & R_n - \frac{\epsilon_n}{\epsilon^\prime_n}R^\prime_n \\
              & = & (686.8\pm83.9) - (656.0\pm49.3) \\
	      & = & 30.8\pm97.3
\end{eqnarray*}
Thus, an upper limit of $\Delta R_n < 180.6$ per year at 90\% confidence limit 
is obtained using a standard, Bayesian prescription
(which is also in good agreement with frequentist prescriptions) \cite{PDG}. 

From \cite{Ejiri}, a vanishing neutron from the $^{16}$O nucleus
results in an excited state which has a branching ratio of 44\% for
producing a 6.18-MeV $\gamma$ and 2\% for a 7.03-MeV $\gamma$.
For a vanishing proton, the distribution is nearly the same,
with a branching ratio of 41\% for a 6.32-MeV $\gamma$ and 4\% for
a 7.0-MeV $\gamma$. The signal extraction procedures previously used
for solar neutrino analyses were applied to simulated nuclear $\gamma$-ray
lines of these energies, combined with a simulated solar neutrino signal. 
The numbers of additional NC events extracted relative to the actual NC 
signals generated were then expressed as fractions of the generated nuclear 
$\gamma$-ray signals. The values of $f_{\gamma n}$ and $f^\prime_{\gamma n}$ 
were then determined by the appropriate weighting of these fractions
in accordance with the relative branching ratios given above.
For Phase I data, it was found that $f_{\gamma n} = 0.99^{+0.01}_{-0.02} $ for both
neutron and proton decay modes. This is as expected since the neutron 
signal in pure D$_2$O results from a 6.25-MeV $\gamma$-ray, which is virtually 
indistinguishable from either 6.18 MeV or 6.32 MeV within the energy resolution
of the detector. The distributions are, therefore, nearly 100\% covariant.
For Phase II data, $f^\prime_{\gamma n} = 0.75^{+0.01}_{-0.01}$ (again, nearly identical
for either decay mode). Once more, this is roughly what is expected
given the additional isotropy information. The lower value of $f^\prime_{\gamma n}$
derived reflects a compromise within the fitting procedure between 
providing a good description of the isotropy distribution and the
energy spectrum expected for neutrons. These same simulated
nuclear $\gamma$-ray lines were also used to determine $\epsilon_\gamma$ and
$\epsilon^\prime_\gamma$. For neutron (proton) decay modes, these were found 
to be $0.51\pm0.01$ ($0.59\pm0.01$) and $0.361\pm0.005$ ($0.425\pm0.006$), 
respectively. 

Thus, an upper limit can be deduced for the number of decay 
$\gamma$-rays at greater than 90\% confidence level of 
$R_\gamma^{\rm lim} < 443$ per year for neutron decay and
$R_\gamma^{\rm lim} < 385$ per year for proton decay.
An upper bound to invisible modes 
of nucleon decay can now be established as follows:

\[ \tau_{\rm inv} > \frac{N_{np}}{R_\gamma^{lim}} \varepsilon_{\gamma} \]

\noindent where $N_{np}$ is the number of neutrons or protons (depending on decay mode)
within the D$_2$O fiducial volume which are bound in $^{16}$O ($1.85\times10^{32}$),
and $\varepsilon_{\gamma}$ is the efficiency for the decay to result in the 
release of a 6 or 7-MeV $\gamma$-ray (0.46 for neutron modes and 0.45 for 
proton modes). Therefore, the comparison of Phase I and Phase II data
from SNO implies that, at greater than 90\% confidence level,

\[\mbox{\it for neutron modes: } \ \ \tau_{\rm inv} >  1.9\times10^{29} \ \mbox{ years.} \]

\[\mbox{\it for proton modes: } \ \ \tau_{\rm inv} >  2.1\times10^{29} \ \mbox{ years.} \]

Prior to this paper, the best constraint on $ n \rightarrow 3 \nu$
used by the Particle Data Group \cite{PDG}
was based on Kamiokande data in which higher energy, but much weaker, 
branches of the de-excitation of the oxygen nucleus were considered and 
yielded a limit of $\tau > 5\times 10^{26}$ years \cite{Suzuki}. It has been proposed
that a similar analysis could be carried out with data from Super-Kamiokande
and, by making use of the carbon nucleus, possibly even in the KamLAND detector
\cite{Kamyshkov}. It has also been noted that the disappearance of a 
proton from the deuteron in heavy water detectors would result in a free neutron, 
which could then be captured to yield a detectable signal for
invisible proton decay (see, for example \cite{Tretyak}). This has 
already been used to yield a lower bound to the proton lifetime in excess of 
$10^{28}$~years for such modes [6,12].
Lead perchlorate has also been suggested as a possible
future detector medium to search for invisible nucleon decay, making use
of de-excitation of the nuclear hole that would be left in $^{35}$Cl, with an 
estimated sensitivity
on the order of $10^{30}$~years for a one kilotonne detector [13,14].
Owing to the extremely low background levels in SNO, the principal
branches of the de-excitations for $^{16}$O have been probed here
and have yielded limits which are within a factor of 5 of this level.
Thus, the constraint presented here is about an order of magnitude
more stringent than the recently published limits on invisible proton-decay 
and 400 times more stringent than previous limits on neutron
modes, such as $ n \rightarrow 3 \nu$.

\begin{acknowledgments}
This research was supported by: Canada: NSERC, Industry Canada,
NRC, Northern Ontario Heritage Fund, Inco, AECL, Ontario Power
Generation, HPCVL, CRC; US: Dept. of Energy; UK: PPARC.
We thank the SNO technical staff for their strong contributions.
\end{acknowledgments}

\newpage


\begin{thebibliography}{99}


\bibitem{Pati} J. Pati, A. Salam and U. Sarkar, {\it Phys. Lett.} {\bf B133}, 
330 (1983)

\bibitem{Mohapatra} R.N.~Mohapatra \& A.~Perez-Lorenzana, {Phys.Rev.} 
{\bf D67}, 075015 (2003)

\bibitem{boger} SNO Collaboration, {\it Nucl. Instr. and Meth.} {\bf A449},
172 (2000)

\bibitem{Ejiri} H.~Ejiri, {\it Phys. Rev.} {\bf C48}, 1442 (1993)

\bibitem{Suzuki} Y.~Suzuki {\em et al.}, {\it Phys. Lett.} {\bf B311}, 357 (1993)

\bibitem{nc_prl} SNO Collaboration, {\it Phys. Rev. Lett.} {\bf 89},
011301-1 (2002)

\bibitem{salt_prl} SNO Collaboration, {\it submitted Phys. Rev. Lett.},
nucl-ex/0309004  (2003)

\bibitem{foot} By contrast, the main thrust of \cite{salt_prl} was a measurement
of the total solar neutrino flux {\em independent} of assumptions regarding the 
CC spectral shape.

\bibitem{PDG} K. Hagiwara {\em et al.}, {\it Phys. Rev.} {\bf D66}, 010001 (2002)


\bibitem{Kamyshkov} Y.~Kamyshkov~and~E.~Kolbe,~{\it Phys.~Rev.}~{\bf D67},~076007~(2003)

\bibitem{Tretyak} {V.I. Tretyak and Y.G. Zdesenko, {\it Phys. Lett.} {\bf B505}, 59 (2001)}

\bibitem{Tretyak2} V.I.~Tretyak and Y.G.~Zdesenko, {\it Phys. Lett.} {\bf B553},
135 (2003)

\bibitem{Boyd} R.N.~Boyd {\em et al.}, hep-ph/0307280 (2003)

\bibitem{foot2} In principle, de-excitation of holes left in $^{35}$Cl can also be explored
in SNO from Phase II data, in which NaCl was added. However, the
concentration of this was only about 0.2\% by weight, so the sensitivity to
nucleon decay is poor relative to $^{16}$O.

\end{thebibliography}

\end{document}